\documentclass[12pt]{iopart} 
\usepackage{iopams}

\usepackage[utf8]{inputenc}
\usepackage{graphicx}

\begin{document}

\title[Ordered structures of tetrahedral patchy particles] {Competing
  ordered structures formed by particles with a regular tetrahedral
  patch decoration}

\author{G\"unther Doppelbauer$^{1,\star}$, Eva G. Noya$^{2,\star}$,
  Emanuela Bianchi$^1$, and Gerhard Kahl$^1$}

\address{$^1$Institut für Theoretische Physik and Center for
  Computational Materials Science (CMS), Technische Universität Wien,
  Wiedner Hauptstra{\ss}e 8-10, A-1040 Wien, Austria \\ $^2$Instituto
  de Qu{\' \i}mica F{\' \i}sica Rocasolano, CSIC, Calle Serrano 119,
  E-28006 Madrid, Spain\\
  $^{\star}$ authors who contributed equally to this work} 

\ead{guenther.doppelbauer@tuwien.ac.at}

Date: \today

\begin{abstract}
We study the ordered equilibrium structures of patchy particles where
the patches are located on the surface of the colloid such that they
form a regular tetrahedron. Using optimization
techniques based on ideas of evolutionary algorithms we identify
possible candidate structures. We retain not only the energetically
most favourable lattices but also include a few energetically less
favourable particle arrangements (i.e., local minima on the enthalpy 
landscape). Using suitably developed Monte Carlo
based simulations techniques in an NPT ensemble we evaluate the
thermodynamic properties of these candidate structures along selected
isobars and isotherms and identify thereby the respective ranges of
stability. We demonstrate on a {\it quantitative} level that the
equilibrium structures at a given state point result from a delicate
compromise between entropy, energy (i.e., the lattice sum) and
packing.
\end{abstract}

\submitto{\JPCM}

\maketitle

\section{Introduction}
\label{sec:introduction}

Patchy particles are colloidal entities whose surfaces are decorated
by well-defined regions, which differ in their interaction behaviour
significantly from the one of the ``naked'' colloidal surface (for an
experimental and theoretical overview see \cite{Patchy_revExp} and
\cite{Patchy_rev2011}, respectively). Since these regions can be
positioned with high accuracy on the particle surface and the spatial extent of
the patches can be tuned in suitable chemical or physical synthesis
processes, the highly directional and selective interactions of patchy
particles have promoted them as very promising entities that are able
to self-assemble into larger target units with desired physical
properties.

The central problem that has to be overcome to achieve this goal is to
acquire a profound understanding of the self-assembly strategies of
patchy particles characterized by a particular patch
decoration. During past years, considerable effort has been dedicated
to solve this intricate and challenging problem via different
numerical and methodological routes
\cite{ZhangKCG05,WilberDLNMW07,NoyaVDL10,RomanoS11,RomanoSS11}. In 
another contribution \cite{patchy3d} we have demonstrated that a suitable combination
of two complementary numerical approaches is able to provide a highly
satisfactory answer to this yet open issue. To be more specific, we
have combined the following two methods: (i) on one hand, an
optimization technique, which employs ideas of evolutionary algorithms
\cite{JPCM10} that is able to predict efficiently and with high
reliability ordered equilibrium structures at {\it vanishing}
temperature; (ii) on the other hand, suitably developed Monte Carlo
simulations \cite{FrenkelLadd,VegaN07} which allow to evaluate accurately, via
thermodynamic integration, the thermodynamic properties of a 
particular ordered structure formed by patchy particles at {\it
  finite} temperature. The most favourable structures with respect to the enthalpy
identified in the first step are considered in the second step as
candidate configurations at finite temperature. The combination of the
two complementary approaches compensates thus for the respective
limitations of the two methods: (i) the risk to lose candidate
structures in a preselection process is now suppressed due to the
systematic search of candidate structures performed in the
optimization step; and (ii) the evaluation of the thermodynamic properties of the systems at finite temperature is now guaranteed by highly reliable and accurate simulations. 

In this contribution we focus on a system of patchy particles where
the patches are positioned in such a way that they form a regular
tetrahedron on the colloidal surface. The ordered equilibrium
structures at zero temperature and the (pressure vs. temperature)
phase diagram obtained via the combined approach outlined above have
been shown and discussed in \cite{patchy3d}. Several aspects that
could not be presented in this rather concise presentation are
discussed in detail in the present contribution. In particular, we
will focus in the following not only on the energetically most
favourable structures as they are suggested at vanishing temperature
by the optimization approach (corresponding to {\it global} minima in
enthalpy at a given pressure value), but also consider those lattices
that represent {\it local} minima, which differ from the optimal solution
by a few percent. For all these structures we have evaluated in a
subsequent step their thermodynamic properties via our simulation
based thermodynamic integration scheme along selected isobars and
isotherms. From this detailed analysis we can obtain a deeper insight
into the complex competition between entropy, energy (i.e., the
lattice sum), and packing which finally defines for a given state
point the energetically most favourable ordered structure.

The paper is organized as follows: in the subsequent section we
briefly present the patchy particle model and give a short summary of
the methods used in this contribution. Results are summarized and
discussed in section 3 and the paper is closed with concluding
remarks.

\section{Model and theoretical tools}
\label{sec:model_theory}

\subsection{The model}
\label{subsec:model}

We use a standard model for patchy particles, that has been proposed by Doye {\it et al.} \cite{DoyeLLANWKL07} and
has been employed in a number of contributions by different
researchers (e.g., see \cite{NoyaVDL07,NoyaVDL10,WilliamsonWDL11,vanderLindenDL11,JPCM10,JPCM11}). The
interaction between two patchy particles is based on a spherical
Lennard-Jones potential with the usual parameters $\epsilon$ and $\sigma$, which specify energy and length units; for interparticle distances $r > \sigma$, this interaction is multiplied by a factor $0 \le V_{\rm ang} \le 1$, which depends on the relative orientations of the particles. Thus, the repulsive part of the Lennard-Jones potential models the spherical colloids, while its attractive part multiplied by $V_{\rm ang}$ models the attraction between the patches.

The spatial extent of the patches is
characterized by a parameter, for which we have
chosen the same value as in \cite{NoyaVDL10,patchy3d}.

\subsection{Theoretical tools}
\label{subsec:tools}

To identify possible candidate structures for the phase diagram at
{\it vanishing} temperature we have used optimization strategies
based on ideas of evolutionary algorithms. 
Such algorithms incorporate concepts of Darwinian evolution in order to tackle optimization problems, i.e., finding extremal values of a cost function $f=f(\mathbf{x})$, where $\mathbf{x}$ is a vector in the parameter search space. A pool of candidate solutions is treated as a ``population'', which undergoes an evolutionary process, including the following operations: selection (i.e., candidate solutions with lower cost function values are more likely to survive within the population and reproduce), reproduction (i.e., features of existing candidate solutions are recombined in order to produce new candidates) and mutation (i.e., features of candidate solutions are randomly modified). We are using a so-called phenotype implementation of an evolutionary algorithm, which combines the aforementioned global optimization steps with local ones, using a limited memory algorithm for bound constrained optimization, relying on the computation of first derivatives of the cost function $\partial f / \partial \mathbf{x}$ \cite{ZhuBN97}. For details we refer to \cite{GottwaldKL05, JPCM10,PauschenweinK08,KahnWK10}.

Our investigations have been carried out in an isobaric-isothermal ensemble.
Thus, the cost function is the Gibbs free energy $G$, which, at vanishing temperature, reduces to the enthalpy $H$, given by $H = U + PV$, $U$
being the lattice sum, $P$ the pressure, and $V$ the volume of the
system. The vectors $\mathbf{x}$ in search space collect the lattice parameters and the coordinates of the particles within the primitive cell. The particle number being $N$, 
we define dimensionless quantities by introducing the packing
fraction $\eta=(\pi/6) (N \sigma^3 / V)$, the reduced enthalpy
$H^{\star}=H/(N \epsilon)$, the reduced energy $U^{\star}=U/(N
\epsilon)$ and the reduced pressure $P^{\star}=P \sigma^3 /\epsilon$;
thus, $H^{\star}=U^{\star}+(\pi/6) (P^{\star}/\eta)$. At some fixed pressure value, we
record during the evolutionary process not only the lattice corresponding to the global enthalpy minimum, but also ordered structures that represent
low-lying local minima on the enthalpy landscape
(differing typically by less than twenty percent from the
energetically most favourable lattice).

For these candidate structures the thermodynamic properties at {\it
 finite} temperature (measured in the dimensionless quantity
$T^\star=k_{\rm B}T/\epsilon$) have then been calculated in
simulations via thermodynamic integration. For this task we have used
a suitably adapted Monte Carlo code, described in detail in
\cite{VegaN07,NoyaVDL10}. 
Briefly, we calculated
the Helmholtz free energy ($A$) for each solid phase at a given thermodynamic state
using the Einstein molecule method \cite{NoyaVDL10}.
In this method the free energy is calculated by thermodynamic integration
using as a reference an Einstein crystal 
with the same structure as the solid
of interest, in which the orientation of the particles is imposed
by an orientational field. For this crystal, the free energy can be numerically
evaluated \cite{NoyaVDL10}.
Once the Helmholtz free energy of a given phase
is known at one thermodynamic state,
its value can be calculated at any other thermodynamic state 
by thermodynamic integration. 
The stability of the structures at finite
$T^\star$ is governed by the minimization of the chemical potential which is given by:
\begin{equation*}
\mu/k_BT = G/Nk_BT =A/Nk_BT + PV/Nk_BT,
\end{equation*}
or, in reduced units, 
\begin{equation*}
\mu^\star/T^\star=A^\star/T^\star+P^\star V^\star/T^\star,
\end{equation*}
where $\mu^\star=\mu/\epsilon$, $A^\star=A/(N\epsilon)$ and
$V^\star=V/(N\sigma^3)$).
When comparing the thermodynamic properties of two competing structures, the relevant quantity is the difference in the chemical potential, $\Delta \mu^\star$. Thus, the dependence of the absolute $\mu^\star$ values on the chosen reference structure does not influence the phase diagram, as long as the same reference is used for all phases.

\section{Results}
\label{sec:Results}

\subsection{Structural variations along isobars}

Figures \ref{fig:minima_g7.298_p025} and \ref{fig:minima_g7.298_p045}
visualize the ordered structures that correspond to the lowest-lying
local enthalpy minima for $P^\star = 2.5$ and $P^\star = 4.5$,
respectively (these values where chosen as particularly interesting based on the knowledge of the $(P^\star,T^\star)$-phase diagram \cite{patchy3d}). The top panel shows, along with the respective enthalpy
values, the two relevant contributions to this quantity, namely the
lattice sum and the packing fractions as bars. These quantities are
shown in relative units of the corresponding values of the energetically
most favourable lattice structures: thus, $U_{\rm i}/U_{\rm opt} > 1$
($< 1$) - where the index ``i'' stands for any of the considered
structures and the index ``opt'' indicates the most
favourable lattice (i.e., structure ``a'' for $P^\star = 2.5$ and ``f'' for $P^\star = 4.5$) - correspond to a higher (lower) degree of bond
saturation than realized in the optimal particle
configuration. Similarly, $\eta_{\rm i}/\eta_{\rm opt} > 1$ ($< 1$)
indicates a better (worse) packing of particles as compared to the
energetically most favourable lattice. In Figures
\ref{fig:isobar_p2.50} and \ref{fig:isobar_p4.50} the internal
energies and the packing fractions of the structures investigated
are displayed as functions of temperature along the two isobars as
obtained by Monte Carlo simulations.

Before we start the discussion of structural variations along two
selected isobars, we mention two structures, that
have raised quite some interest in the literature
\cite{Romano10,RomanoSS11,NoyaVDL10}, namely diamond cubic and diamond
hexagonal phases (especially the former, which has potential use in the fabrication of materials with photonic band gaps \cite{MaldovanT04}). In our calculations, these structures appear as
corresponding to low-lying minima on the enthalpy landscape for very low pressure
values, but as discussed in \cite{NoyaVDL10}, are never stable for the
values of the potential parameters we are using in this contribution. 

\subsubsection{Isobar at $P^\star = 2.5$}

\begin{figure}[htbp]
\begin{center}
\includegraphics[width=8.0cm]{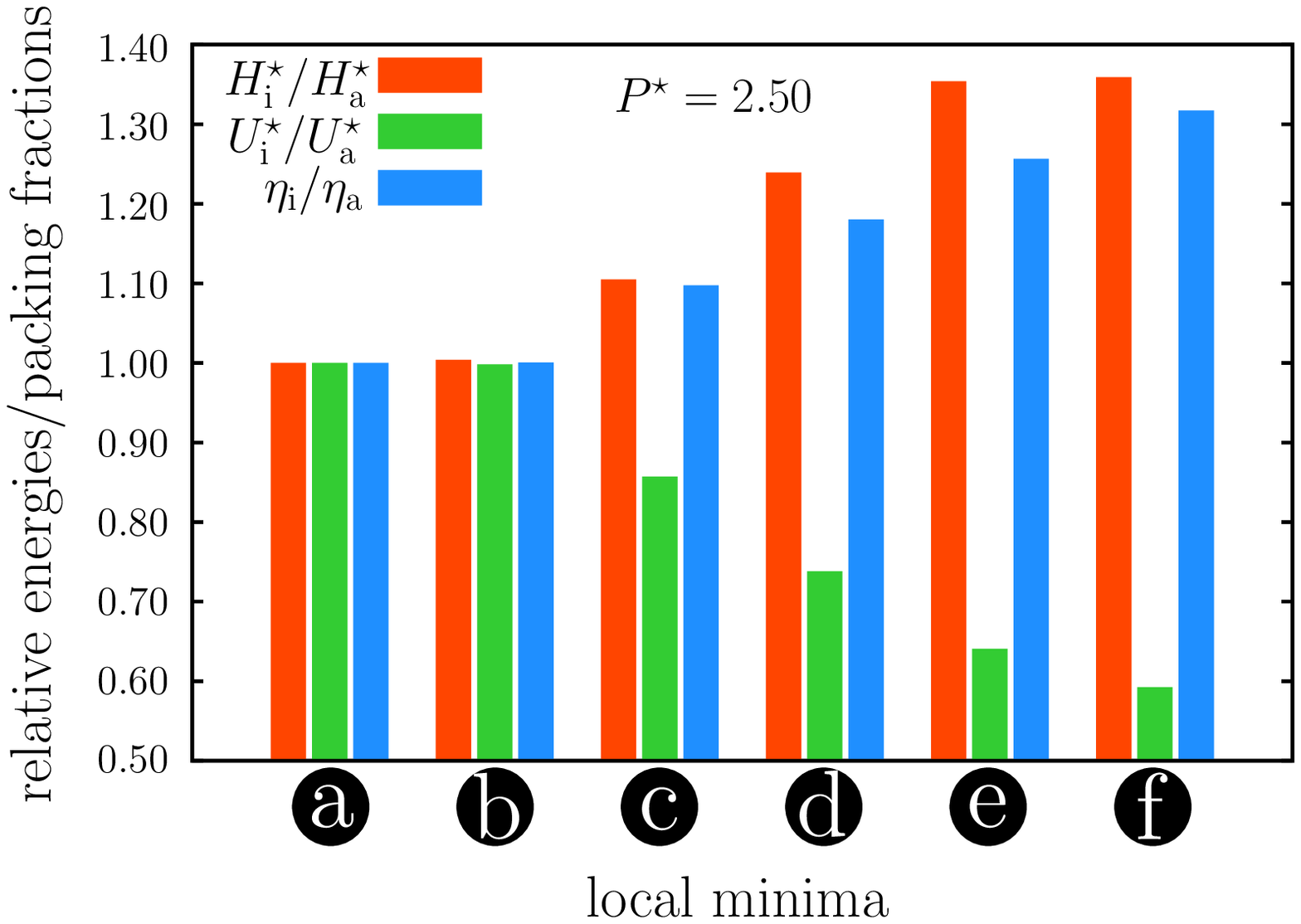}
\end{center}
\begin{center}
\begin{tabular}{r | l}
\parbox[c]{1.8cm}{\includegraphics[width=0.8cm]{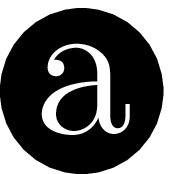}
double diamond/
bcc-like

shifted} 
\parbox[c]{5.2cm}{\includegraphics[width=5.0cm]{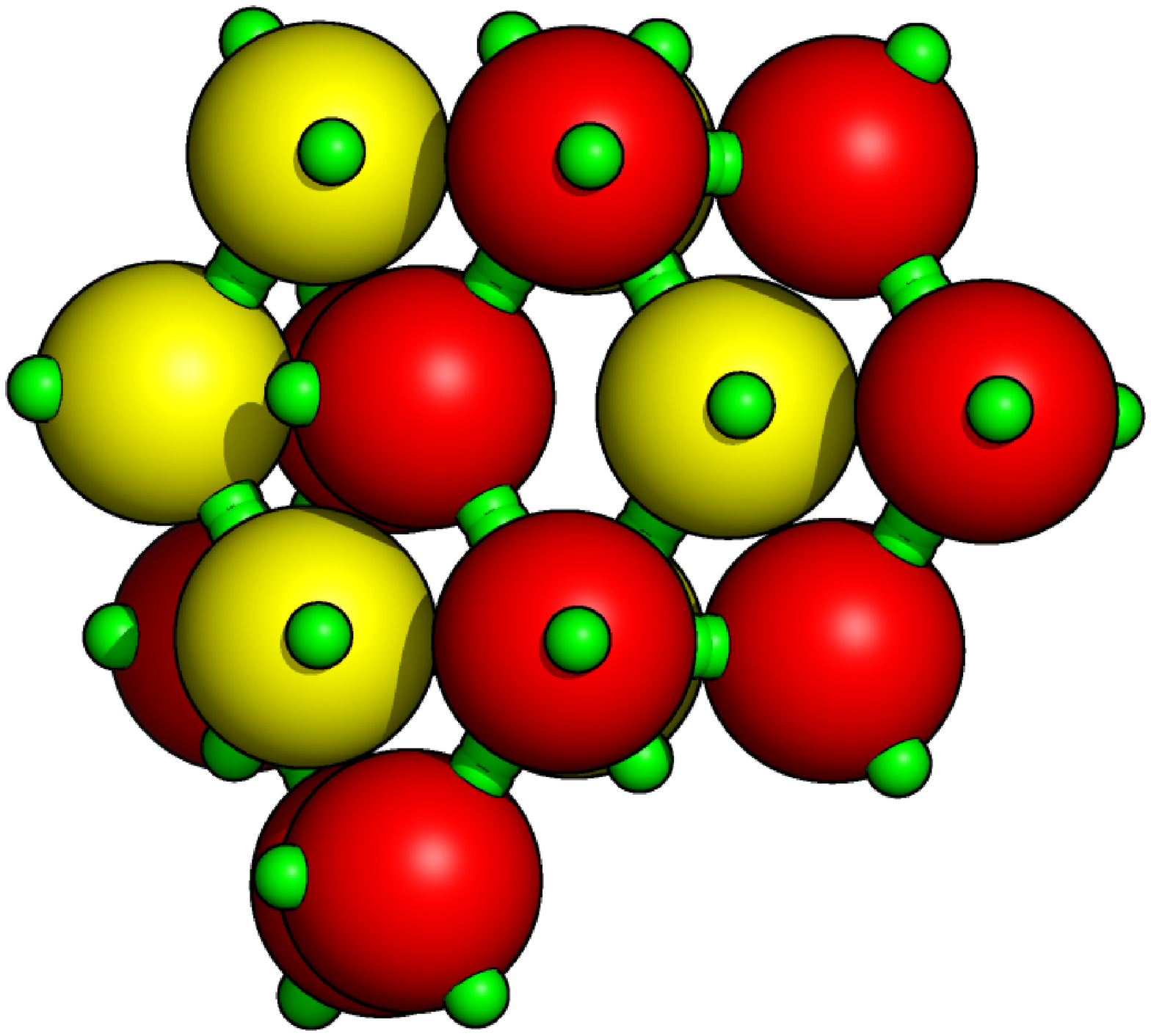}} & 
\parbox[c]{1.8cm}{\includegraphics[width=0.8cm]{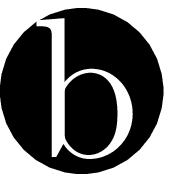}
double diamond/
bcc-like

symmetric} 
\parbox[c]{5.2cm}{\includegraphics[width=5.0cm]{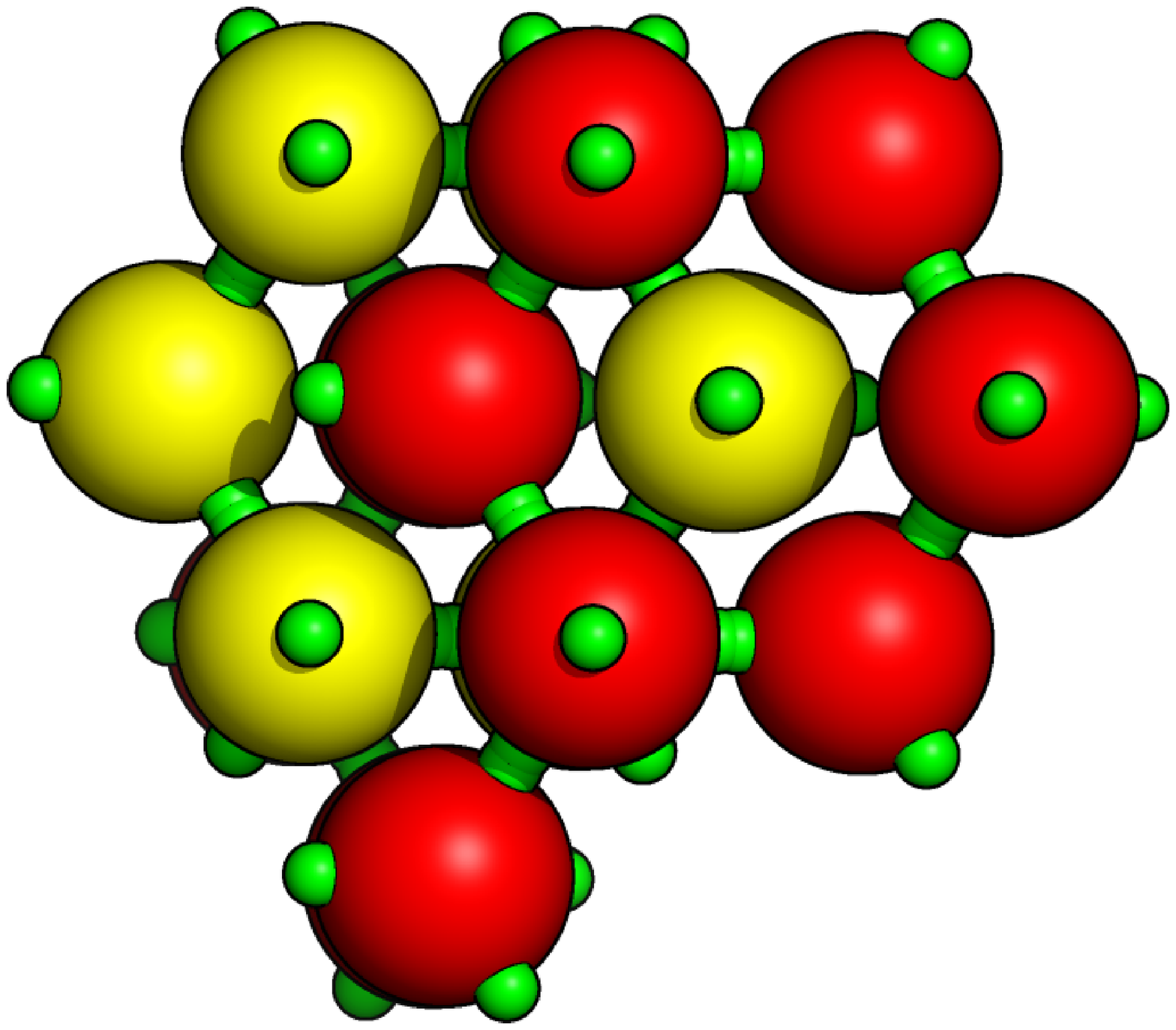}
} \\
\hline
\parbox[c]{1.8cm}{\includegraphics[width=0.8cm]{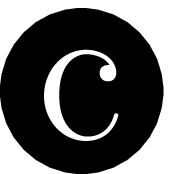}
double diamond/
bcc-like

broken} 
\parbox[c]{5.2cm}{\includegraphics[width=5.0cm]{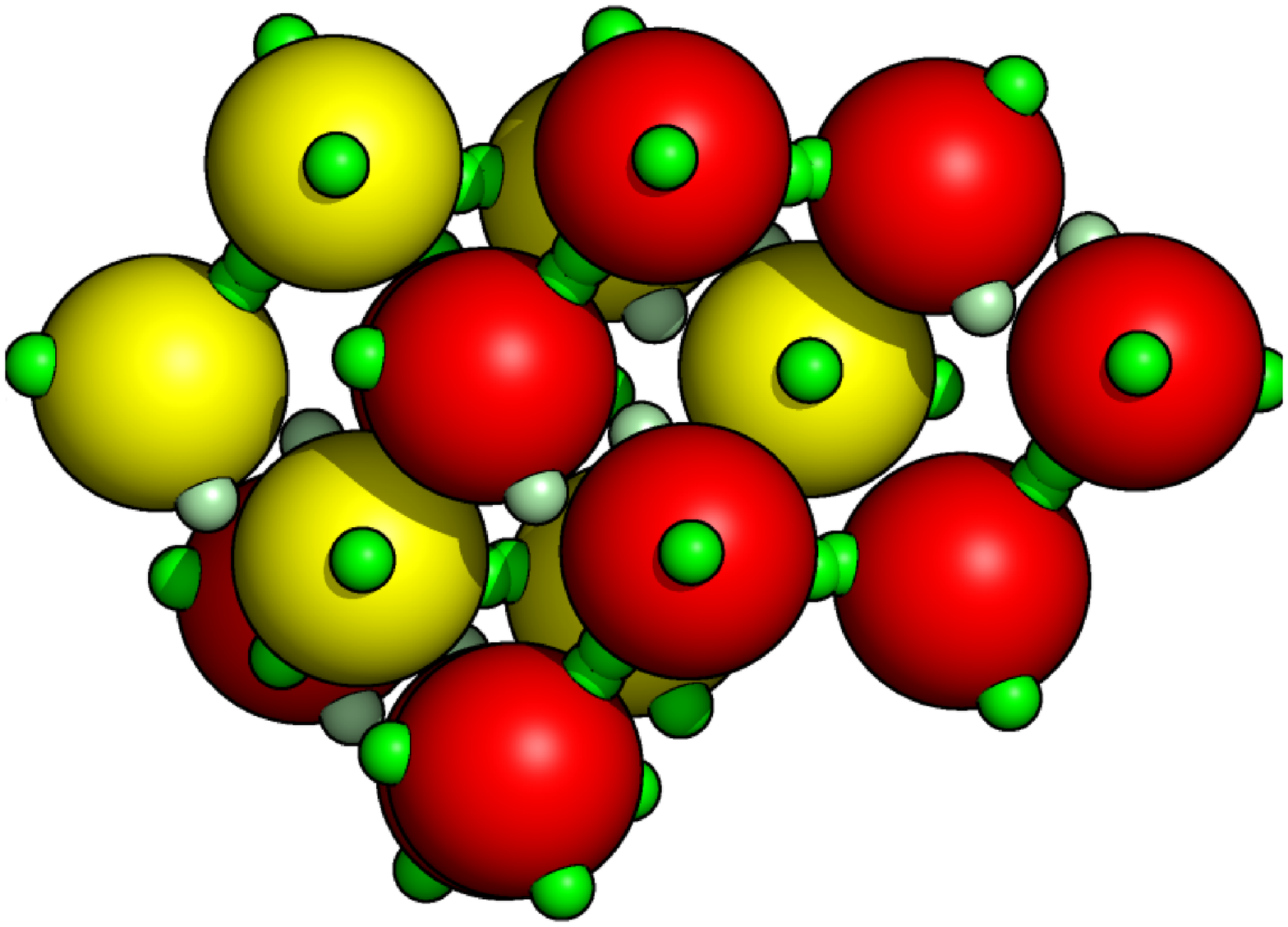}} & 
\parbox[c]{1.8cm}{\includegraphics[width=0.8cm]{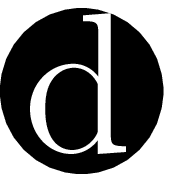}

hexagonal
double
layers} 
\parbox[c]{5.2cm}{\includegraphics[width=5.0cm]{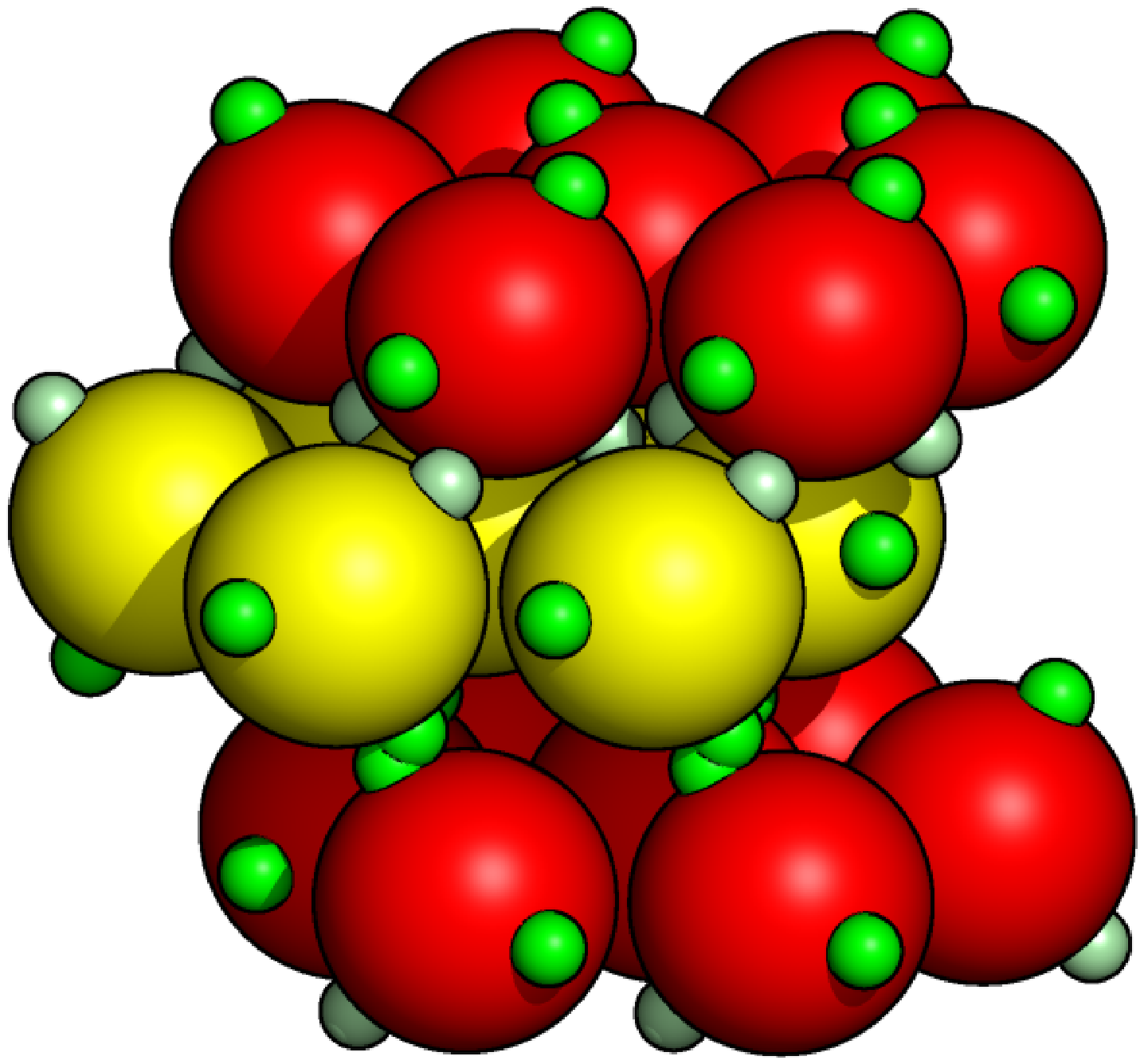}
} \\
\hline
\parbox[c]{1.8cm}{\includegraphics[width=0.8cm]{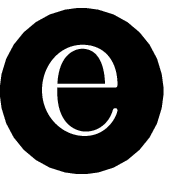}

fcc-like 1} 
\parbox[c]{5.2cm}{\includegraphics[width=5.0cm]{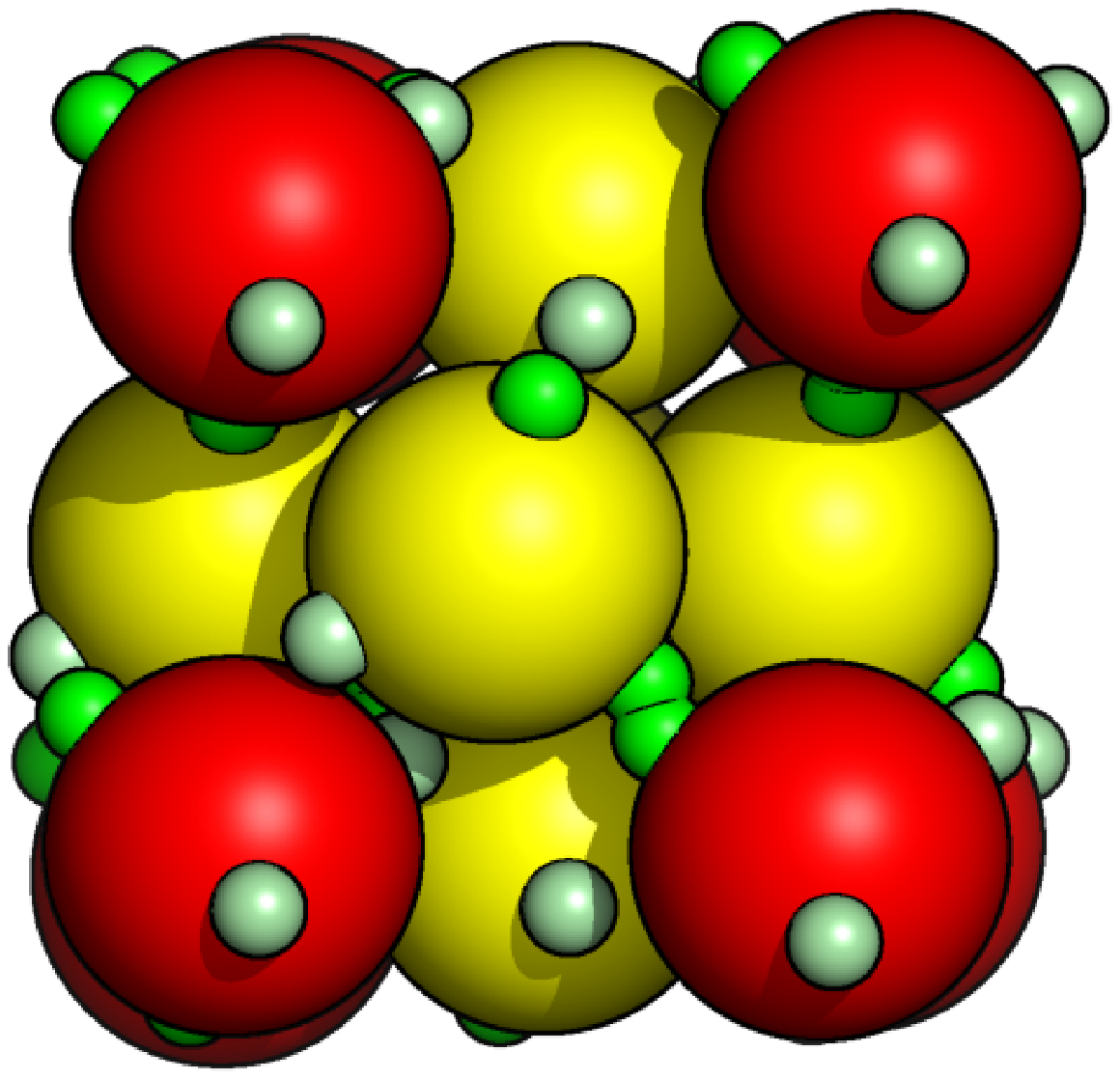}} & 
\parbox[c]{1.8cm}{\includegraphics[width=0.8cm]{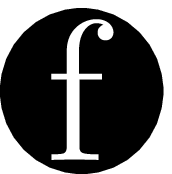}

fcc-like 2} 
\parbox[c]{5.2cm}{\includegraphics[width=5.0cm]{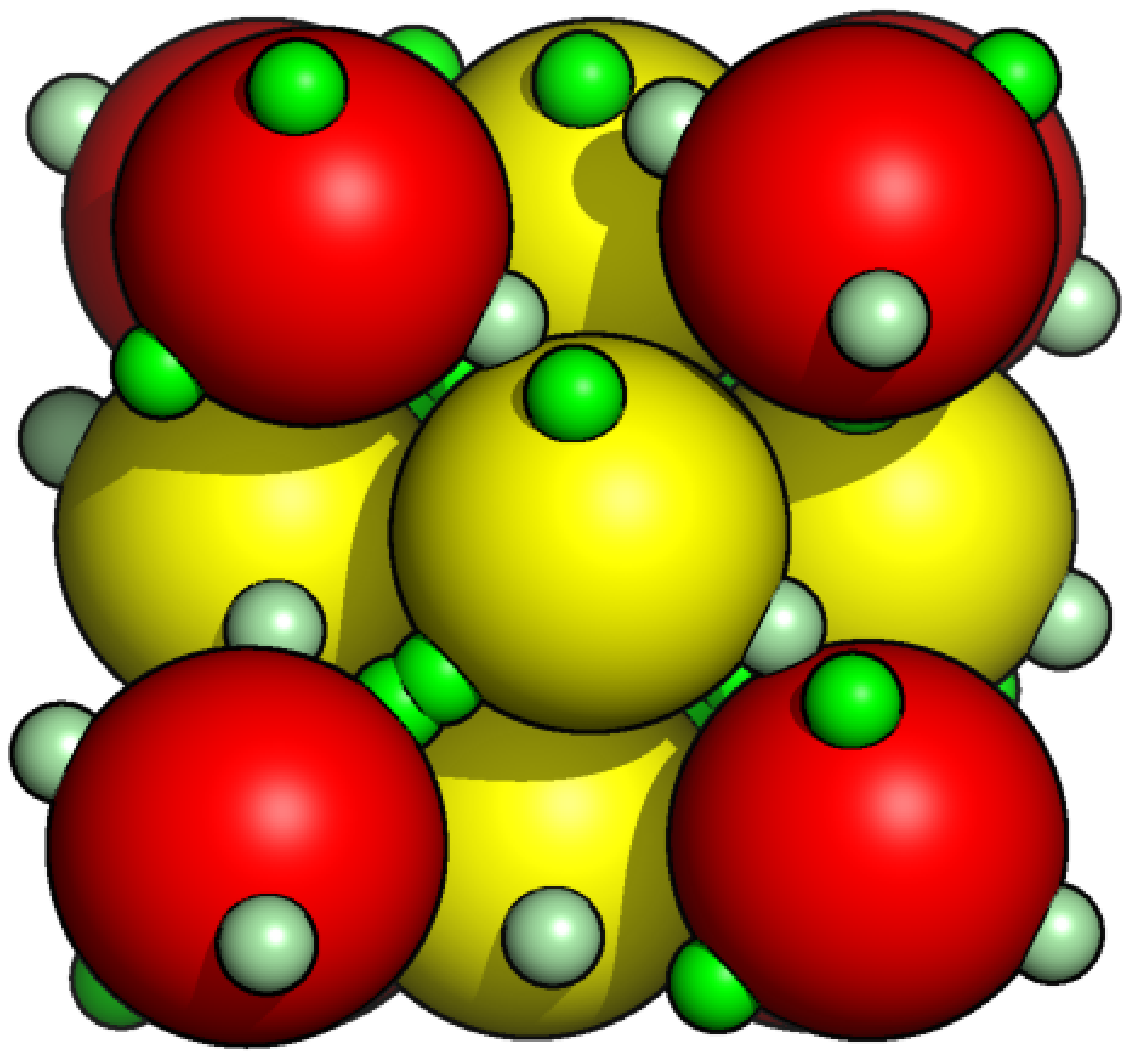}}
\end{tabular}
\end{center}
\caption{(Colour online) Top panel: enthalpies $H^\star_{\rm i}$, binding
  energies (i.e., lattice sums) $U^\star_{\rm i}$, and packing fractions
  $\eta_{\rm i}$ of the six lowest, structurally different local
  enthalpy minima identified by the evolutionary algorithm for
  $P^{\star}=2.50$ with i = ``a'' to ``f''. Values are given in units
  of the respective values of the energetically most favourable
  lattice (``a''). Other panels: visual representations of these
  structures as labeled. Particles are coloured red and yellow as a
  guide to the eye \cite{note1}. Fully bonded patches are coloured
  bright green, weakly- or non-bonded patches are shown in pale
  green. Structures ``a'', ``b'', ``c'', ``e'', and ``f'' are
  thermodynamically stable in certain regions of the
  $(P^{\star},T^{\star})$-phase diagram (cf. \cite{patchy3d}).}
\label{fig:minima_g7.298_p025}
\end{figure}

 The six most favourable structures with respect to the enthalpy,
identified at $P^\star=2.5$ are summarized in Figure
\ref{fig:minima_g7.298_p025} (labeled in this Figure and in the
following by ``a'' to ``f'') along with an analysis of how the
respective values for the enthalpy are split up into the lattice sum
and the packing fraction.

The global enthalpy minimum for this pressure value corresponds to, as
previously reported in \cite{Romano10,NoyaVDL10,patchy3d}, a
bcc-like structure that consists of two interpenetrating, but
virtually non-interacting diamond lattices. This structure emerges in
two different, but closely related versions: the particles of one
diamond sublattice can be located exactly in the centers of the voids
of the six-particle rings making up the other diamond sublattice
(``b'' in Figure \ref{fig:minima_g7.298_p025}) or the positions
of the particles in each sublattice can be slightly shifted against
each other (``a'' in Figure \ref{fig:minima_g7.298_p025}). At
$T^\star=0$, the latter structure is energetically more favourable:
its enthalpy value is by 0.3 percent smaller as compared to the
enthalpy of the former. This can be understood by the fact that the
bond lengths of the shifted configuration are slightly closer to the
ideal value (i.e., the position of the minimum of the Lennard-Jones
potential) at the same packing fraction ($\eta=0.54$) as the symmetric
configuration. Already at a minute value of temperature, these two
structures become indistinguishable (cf. Figure
\ref{fig:isobar_p2.50}).  

Configuration ``c'', which has the third-lowest enthalpy value at
$P^\star=2.5$ and vanishing temperature, is also closely related to the double
diamond configuration: compared to structure ``b'', bonds between
two oppositely located pairs of particles within each six-particle
ring are broken (see pale green patches displayed in panel ``c'' of
Figure \ref{fig:minima_g7.298_p025}); the emerging half-rings are
slightly distorted and displaced with respect to each other. This
results in a higher packing fraction (by 9.8 \%) but a substantial
increase in the lattice sum (by 14.2 \%); in total this leads to an
increase in the enthalpy by 10.5 \% with respect to the energetically
most favourable particle arrangement at $P^\star=2.5$ (``a''). 
We note that at vanishing temperature structure ``c'' becomes more favourable than
  the fully bonded ones at $P^\star=3.2$ (see Figure
  \ref{fig:enthalpy_mu}, top panel). However, at finite temperature, as
entropic effects set in, the relatively low enthalpy value is soon
overruled: our simulation results indicate that the structure is only
stable for temperatures up to $T^\star \simeq 0.006$.

For structure ``d'', corresponding to the fourth-lowest local enthalpy
minimum, the
discrepancy between bonding and packing is even more pronounced: we
identify a relatively dense configuration ($\eta=0.63$),
consisting of hexagonally arranged particles, where each layer is
strongly bonded with one of its neighbouring layers and unconnected
with the other one. To be more specific, each particle forms bonds via
three of its patches: two of these bond with particles
within the same layer, the third one connects the particle to a
neighbouring layer. Our simulations at finite temperature show that
this configuration is never the most stable one in the ($P^\star,
T^\star$)-phase diagram.

The structure representing enthalpy minimum five is a non-close packed
($\eta=0.68$) fcc-like lattice (``e'' in Figure
\ref{fig:minima_g7.298_p025}), where each particle has two saturated
bonds. Within the fcc-picture, the particles located at the vertices
of the cube differ in their orientation from the particles that occupy
the centers of the faces. Despite its high packing fraction, this
structure is never stable at vanishing temperature due to the small
number of saturated bonds. However, a slightly modified version of
this lattice (previously discussed in
\cite{NoyaVDL10}), where each particle is rotated in order to replace a
single, fully saturated bond by two weaker bonds, is found to be
stable at finite temperatures in the ($P^\star, T^\star$)-phase
diagram \cite{patchy3d} for $T^\star \gtrsim 0.06$ over a pressure
range steadily increasing with temperature.

Structure ``f'', corresponding to the sixth local enthalpy minimum, is another fcc-like configuration, which achieves a relatively high value
for the packing fraction, namely $\eta=0.71$ (compared to $\eta=0.74$
for close packed hard spheres). Again, each particle forms two bonds
and particles located at different positions in the cubic cell are
oriented in two different directions. For pressure
values above $P^\star=3.4$, this structure corresponds to the global
enthalpy minimum at $T^\star=0$ and has a broad region of stability at
finite temperature \cite{patchy3d}.

In Figure \ref{fig:isobar_p2.50} we have depicted our MC results for
the internal energy $U^\star$ and for the packing fraction $\eta$
evaluated for the six lattice structures considered at $P^\star = 2.5$
over a representative temperature range. The curves are shown over the
respective ranges of mechanical stability for each ordered structure. Note that
the simulation data for $U^\star$ and $\eta$ can smoothly be
extrapolated for $T \to 0$ to the respective values obtained via the
optimization algorithm which nicely demonstrated the internal
consistency of our combined approach. The $U^\star$- and $\eta$-curves
for lattices labeled ``a'' and ``b'' become essentially
indistinguishably close, even for the smallest finite temperature investigated. In
the displayed temperature range no phase transition takes place
(cf. \cite{patchy3d}).

As expected we observe a monotonous increase of $U^\star$ with
temperature for all structures (``a'' to ``f''), giving
evidence that the bond saturation decreases with increasing
$T^\star$. From the smoothness of the curve one can conclude that no
structural transition takes place in the displayed
temperature interval. Concomitantly the packing fraction decreases --
as expected -- monotonically with increasing temperature. It should be
emphasized that for the energetically most favourable structure(s) (``a''
and ``b'') the packing fraction remains essentially constant within
the observed temperature interval.

\subsubsection{Isobar at $P^\star = 4.5$}

\begin{figure}[htbp]
\begin{center}
\includegraphics[width=8.0cm]{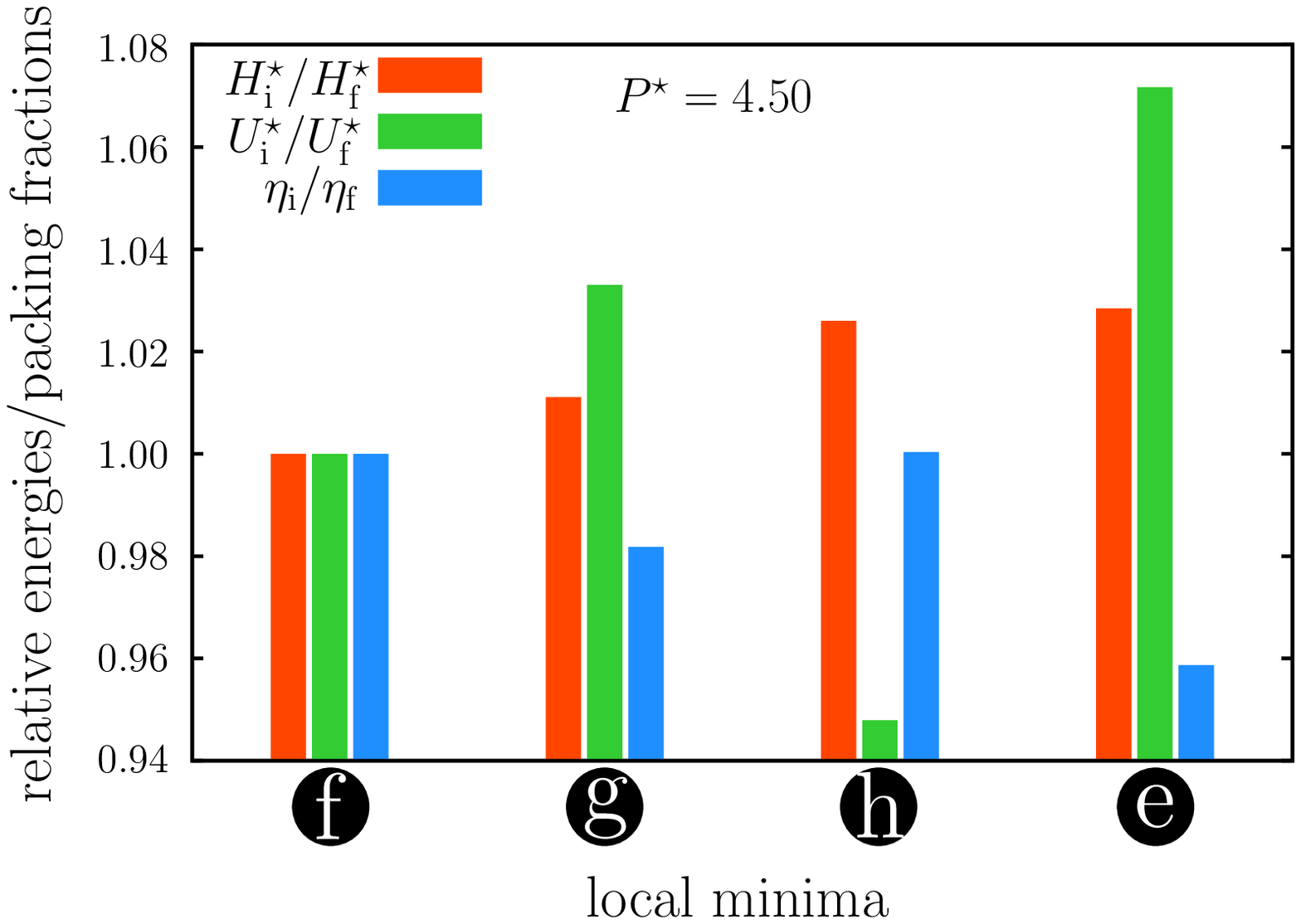}
\end{center}
\begin{center}
\begin{tabular}{r | l}
\parbox[c]{1.8cm}{\includegraphics[width=0.8cm]{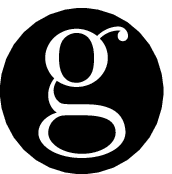}

hcp-like 1} 
\parbox[c]{5.2cm}{\includegraphics[width=5.0cm]{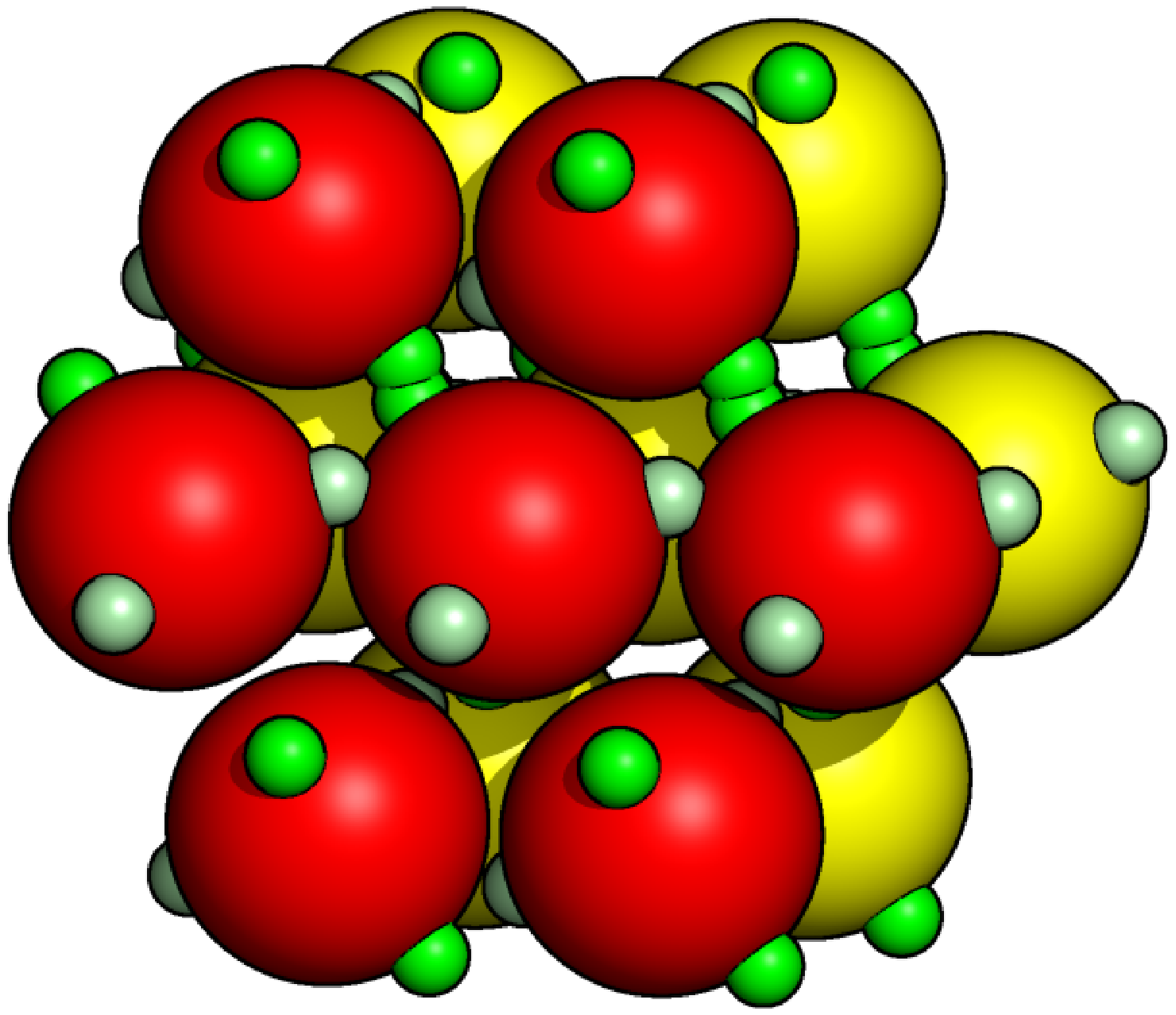}} & 
\parbox[c]{1.8cm}{\includegraphics[width=0.8cm]{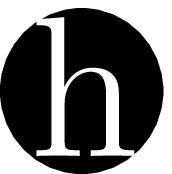}

hcp-like 2} 
\parbox[c]{5.2cm}{\includegraphics[width=5.0cm]{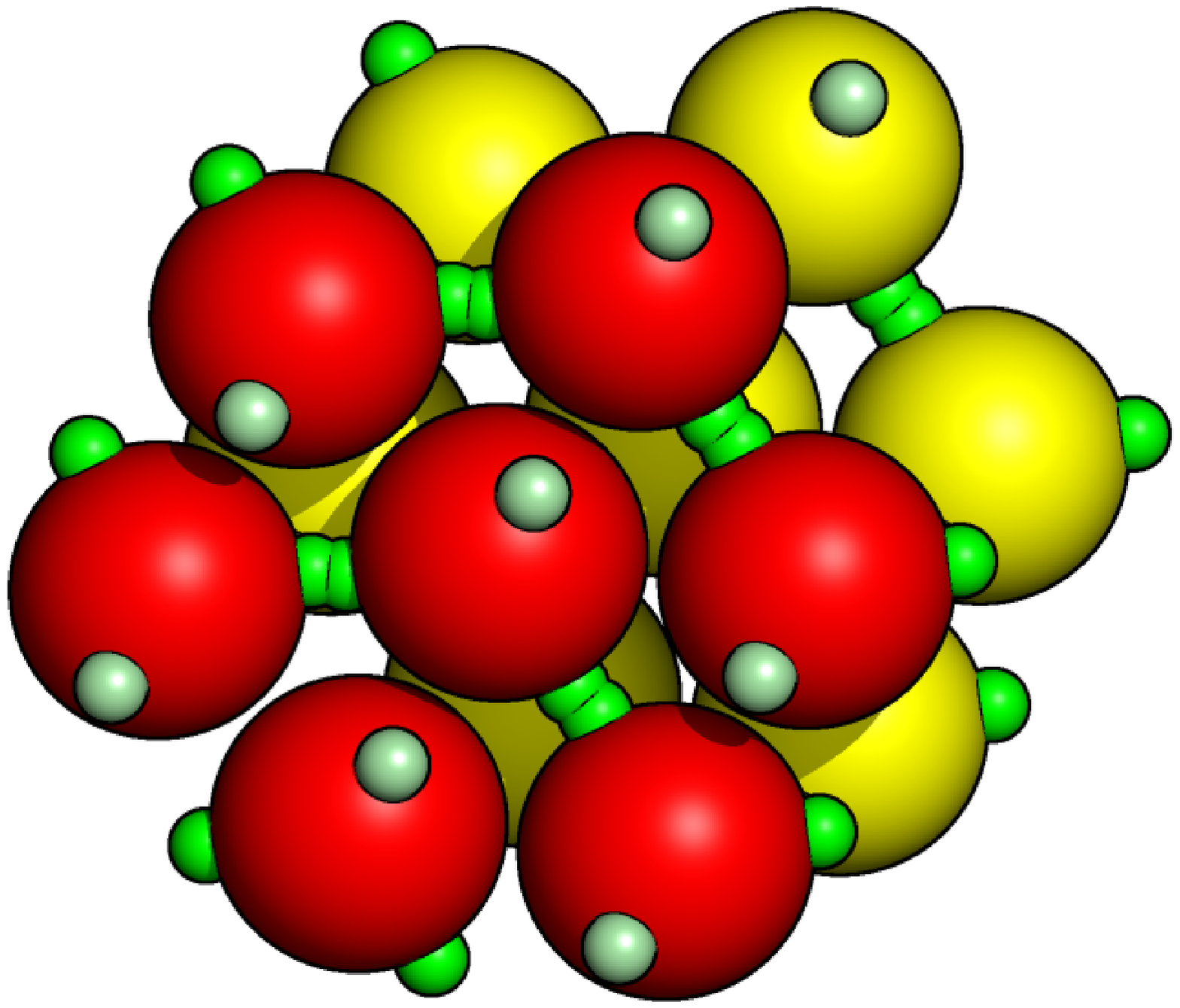}}
\end{tabular}
\end{center}
  \caption{(Colour online) Top panel: enthalpies $H^\star_{\rm i}$, binding
    energies (i.e., lattice sums) $U^\star_{\rm i}$, and packing fractions
    $\eta_{\rm i}$ of the four lowest, structurally different local
    enthalpy minima identified by the evolutionary algorithm for
    $P^{\star}=4.50$ with i = ``e'' to ``h''. Values are given in
    units of the respective values of the energetically most
    favourable lattice (``f''). Other panels: visual
    representations of structures ``g'' and ``h''; the other
    two lattices are displayed in Figure \ref{fig:minima_g7.298_p025}.
    Particles are coloured red and yellow as a guide to the eye
    \cite{note1}. Fully bonded patches are coloured bright green,
    weakly- or non-bonded patches are shown in pale green. Structures ``e'' and ``f'' are thermodynamically stable in certain regions of
    the $(P^{\star},T^{\star})$-phase diagram (cf. \cite{patchy3d}).}
\label{fig:minima_g7.298_p045}
\end{figure}

\begin{figure}[htbp]
\begin{center}
\includegraphics[width=11.cm]{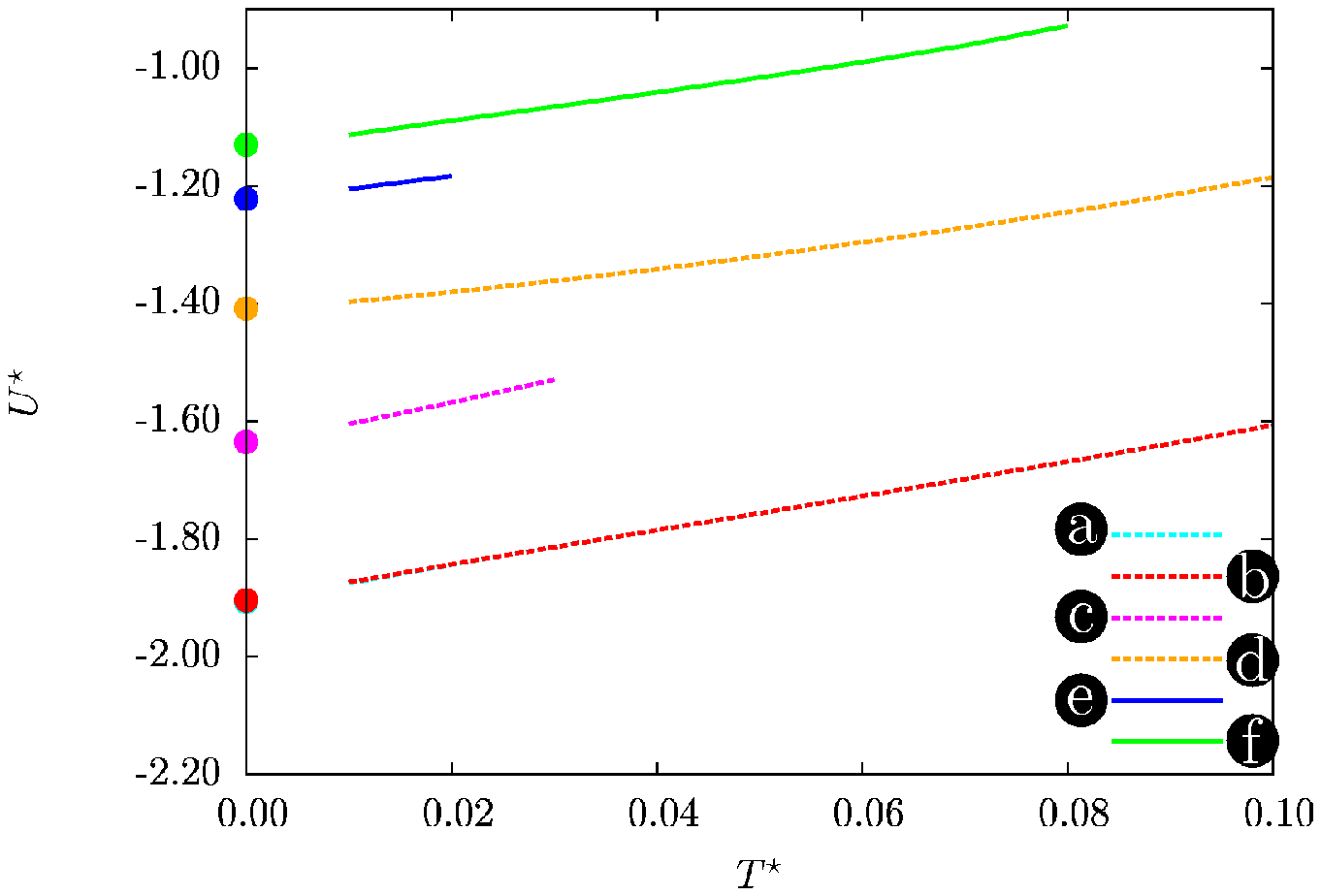}
\includegraphics[width=11.cm]{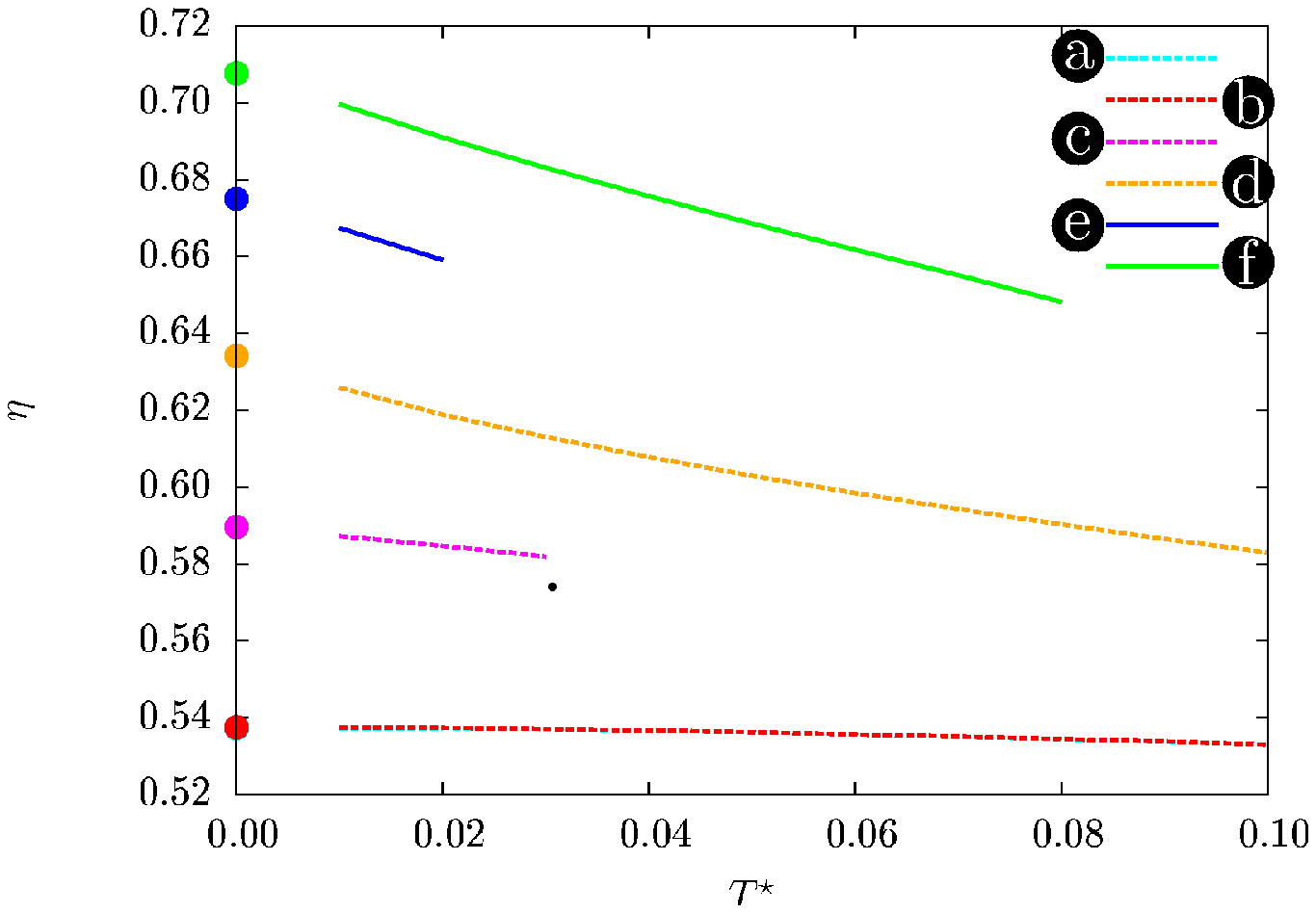}
\end{center}
\caption{(Colour online) Reduced binding energy (i.e., lattice sum)
  $U^{\star}$ (top panel) and packing fraction $\eta$ (bottom panel)
  of six different competing crystal structures as functions of
  temperature $T^\star$, obtained from MC simulations. These lattices have been identified as low-lying local
  enthalpy minima along the isobar $P^\star=2.50$. The lines for the
  double diamond/bcc shifted lattice (``a'') and the double
  diamond/bcc symmetric structure (``b'') coincide within
  line-thickness. Data are shown only over the temperature ranges
  where the respective structures are mechanically stable. Dots on the vertical
  axes (i.e., $T^\star =0$) represent results for $U^\star$ and $\eta$
  obtained from the evolutionary algorithm.}
  \label{fig:isobar_p2.50}
\end{figure}

For this pressure value, we have considered the structures that
correspond to the four lowest lying local enthalpy minima; among those
lattices two (``e'' and ``f'' in Figure
\ref{fig:minima_g7.298_p025}) have already been discussed for the case
$P^\star = 2.5$. The other two structures, labeled ``g'' and ``h'' are
depicted in Figure \ref{fig:minima_g7.298_p045}.

The global minimum in enthalpy at vanishing temperature still
corresponds to the almost close-packed fcc-like structure described
above (``f'' in Figure \ref{fig:minima_g7.298_p025}).  The
second-lowest minimum of the enthalpy landscape corresponds to an
hcp-like structure (``g'' in Figure
\ref{fig:minima_g7.298_p045}). Similar to the packed fcc-like
case, each particle forms two bonds -- one
with a particle within the same layer, the other connecting to a
particle in a neighbouring layer; however, here the bonding angles are found to be considerably
closer to the ideal values (i.e., the patches directly face each
other), resulting in a slightly lower lattice sum
$U^\star$. Nevertheless, the packing fraction of this structure
($\eta=0.70$) is again lower, rendering it, in total, less favourable
than configuration ``f''. 
The third local enthalpy minimum is represented by another hcp-like structure
(``h'' in Figure \ref{fig:minima_g7.298_p045}).
Similar to the
hexagonal configuration mentioned in the previous paragraph (``d'' in Figure \ref{fig:minima_g7.298_p025}), each particle forms two
intra-layer bonds; however, in configuration ``h'', there are no
inter-layer bonds at all. This lattice reaches a packing
fraction similar to the one of the global minimum structure (``f'') (at vanishing and very low $T^\star$ it is even slightly
  higher); however, it has a considerably higher $U^\star$-value,
making it thus unstable in the entire ($P^\star, T^\star$)-phase
diagram. Local minimum four corresponds to the non-close-packed fcc-like
structure already identified at $P^\star=2.5$ (``e'' in Figure
\ref{fig:minima_g7.298_p025}).

\begin{figure}[htbp]
\begin{center}
\includegraphics[width=11.cm]{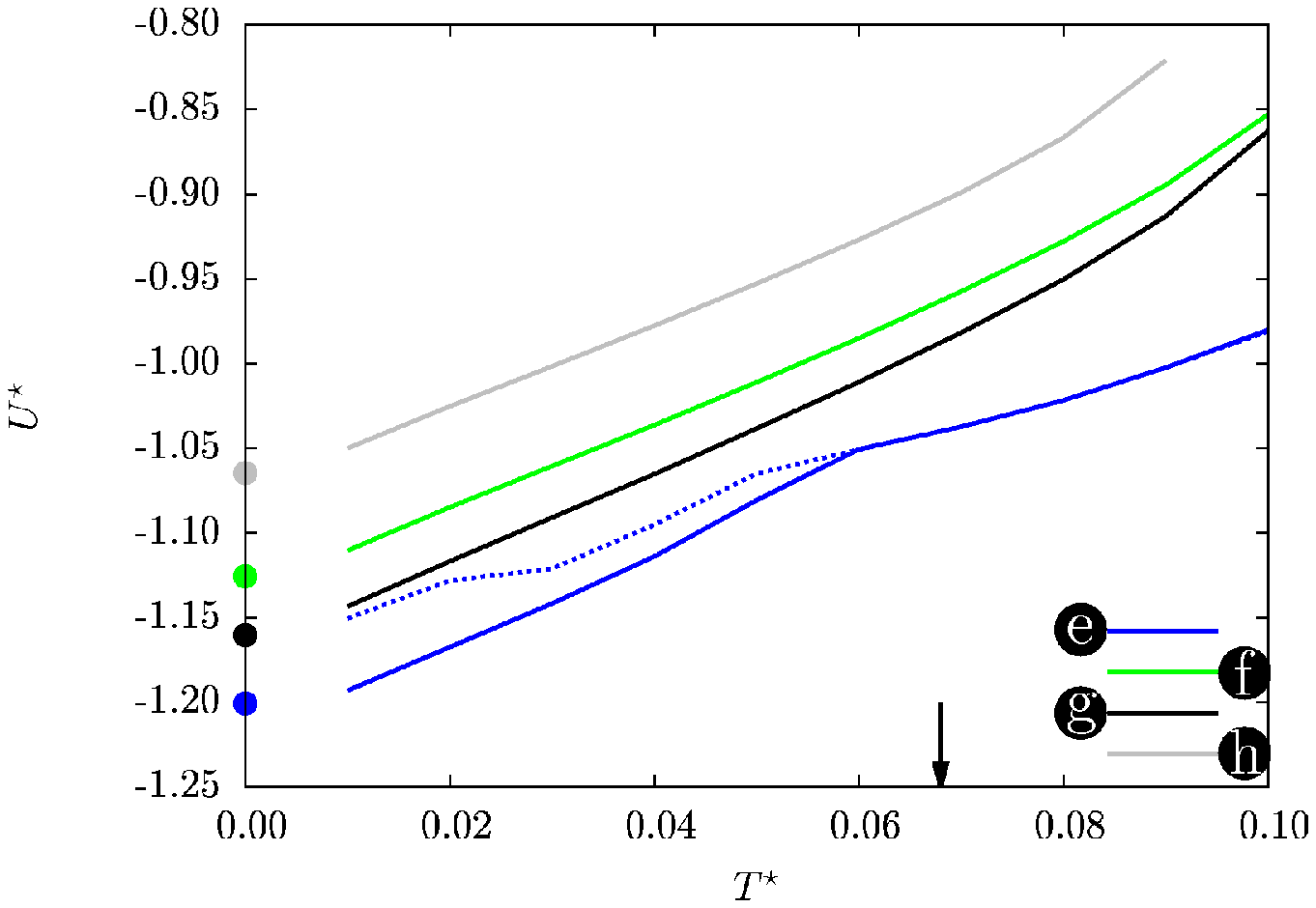}
\includegraphics[width=11.cm]{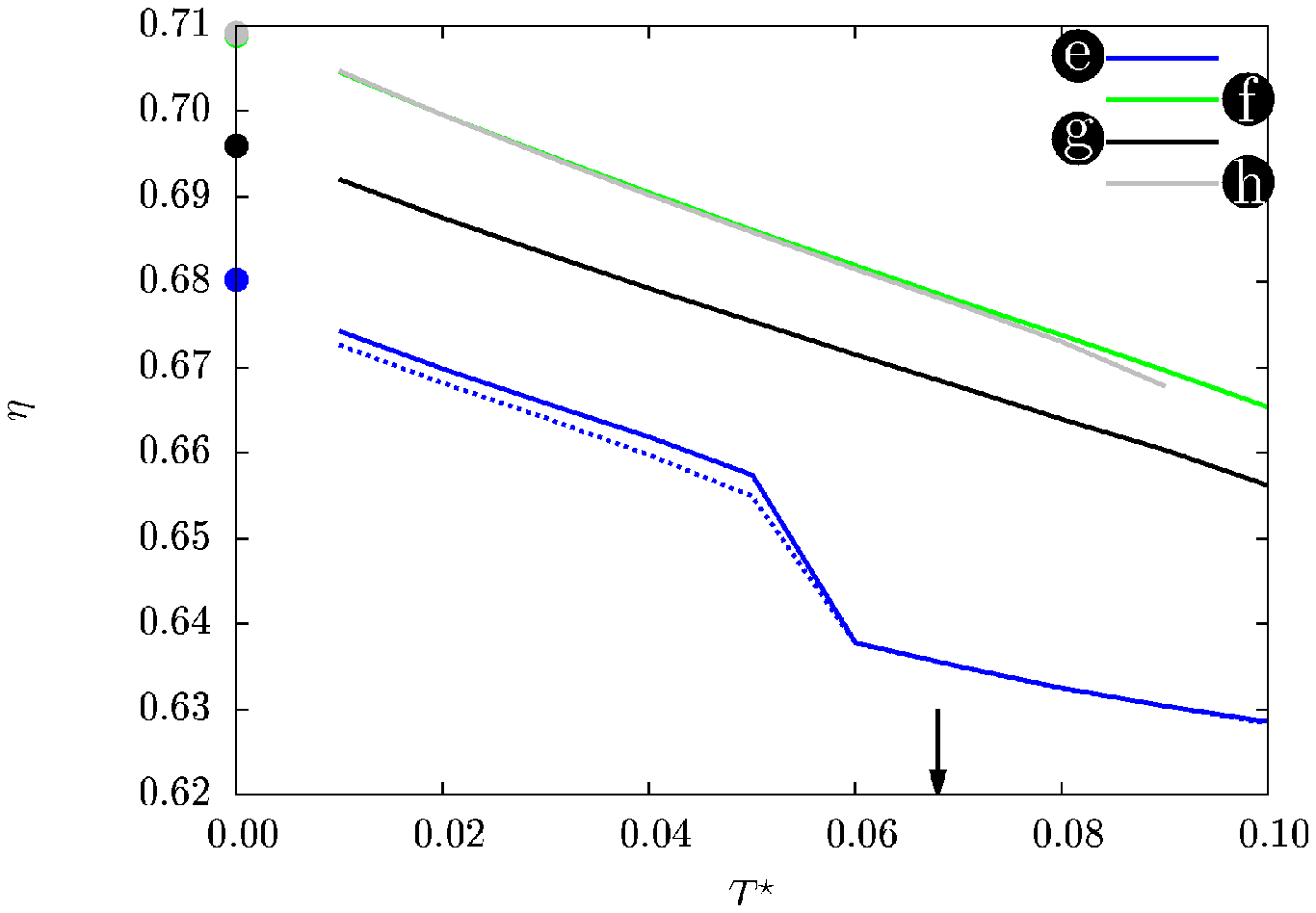}
\end{center}
\caption{(Colour online) Reduced binding energy (i.e., lattice sum)
  $U^{\star}$ (top panel) and packing fraction $\eta$ (bottom panel)
  of four different competing crystal structures as functions of
  temperature $T^\star$, obtained from MC simulations. These lattices have been identified as low-lying local
  enthalpy minima along the isobar $P^\star=4.50$. Data are shown only
  over the temperature ranges where the respective structures are
  mechanically stable. Dots on the vertical axes (i.e., $T^{\star}=0$) represent
  results for $U^\star$ and $\eta$ obtained from the evolutionary
  algorithm. The vertical arrow indicates the temperature values
  where a structural phase transition takes place (i.e., between structures ``e'' and ``f'' at $T^\star=0.068$). For a discussion of the dotted blue lines, we
  refer to the main text.}
\label{fig:isobar_p4.50}
\end{figure}

In Figure \ref{fig:isobar_p4.50} we display $U^\star$ and $\eta$ as
functions of temperature which show -- as expected -- a monotonous
increase and a monotonous decrease with increasing temperature,
respectively. For structure ``e'' at this pressure value, we observe discontinuous changes
both in $U^\star$ and $\eta$ at $T^\star \simeq 0.05$ indicating a
structural change that occurs with increasing temperature. The main
feature of this transition is a reorientation of the particles, so
that one well-aligned patch-patch bond is replaced by two weaker bonds
with less optimal alignment. As mentioned above, the high temperature version of structure ``e''
(i.e., the one with two weaker bonds) has a region of stability in the
phase diagram. 
We note that when cooling this structure down again, we do not
recover structure ``e'' (see the dotted blue lines in Fig \ref{fig:isobar_p4.50}).
Instead at low temperature we obtain a structure with slightly lower 
packing fraction but much higher energy. We have not studied this issue
further, but it is possible that a different behaviour is observed by changing
the speed of quenching.

\subsection{Structural variations along isotherms}

\begin{figure}[htbp]
\begin{center}
\includegraphics[width=11.cm]{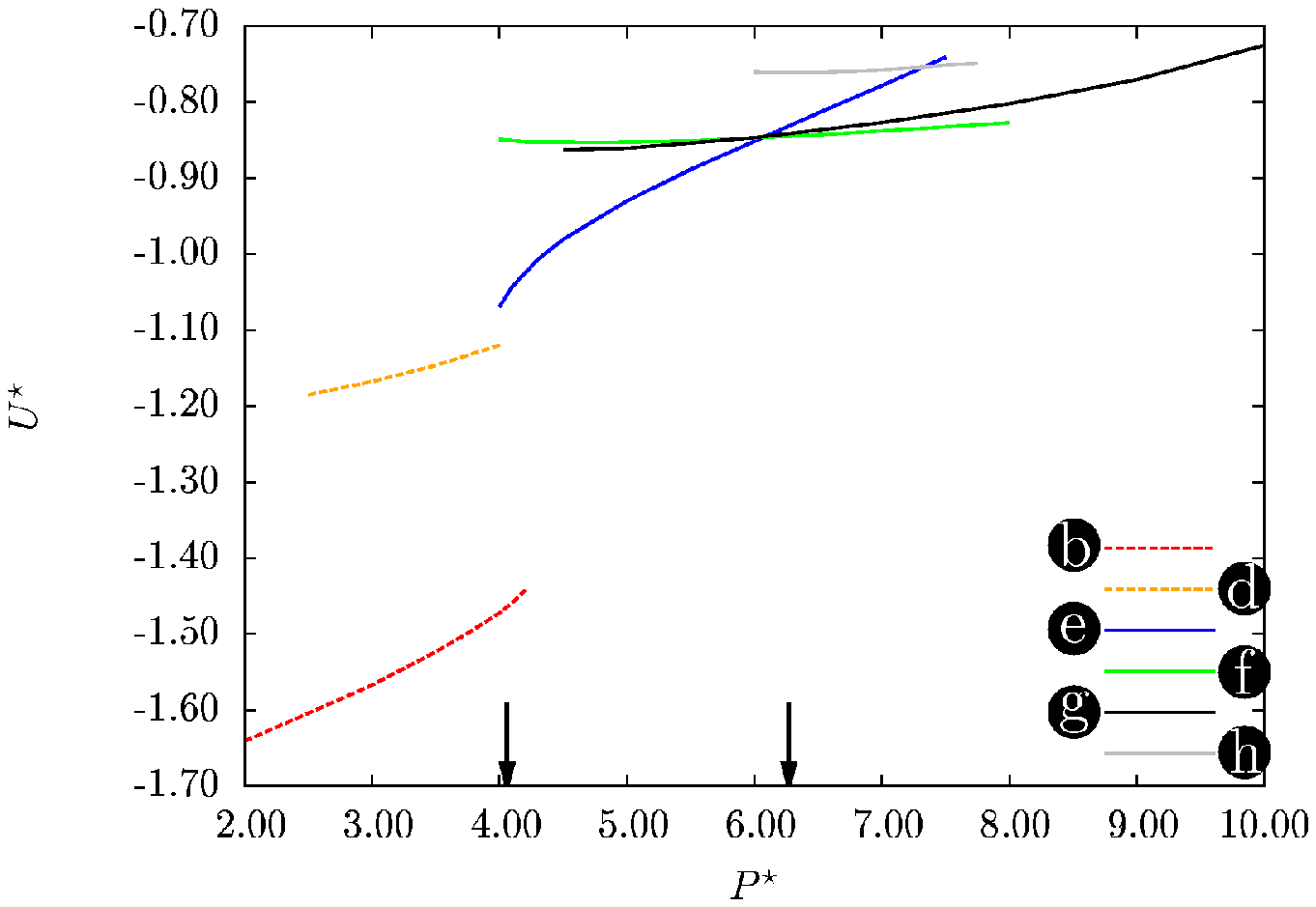}
\includegraphics[width=11.cm]{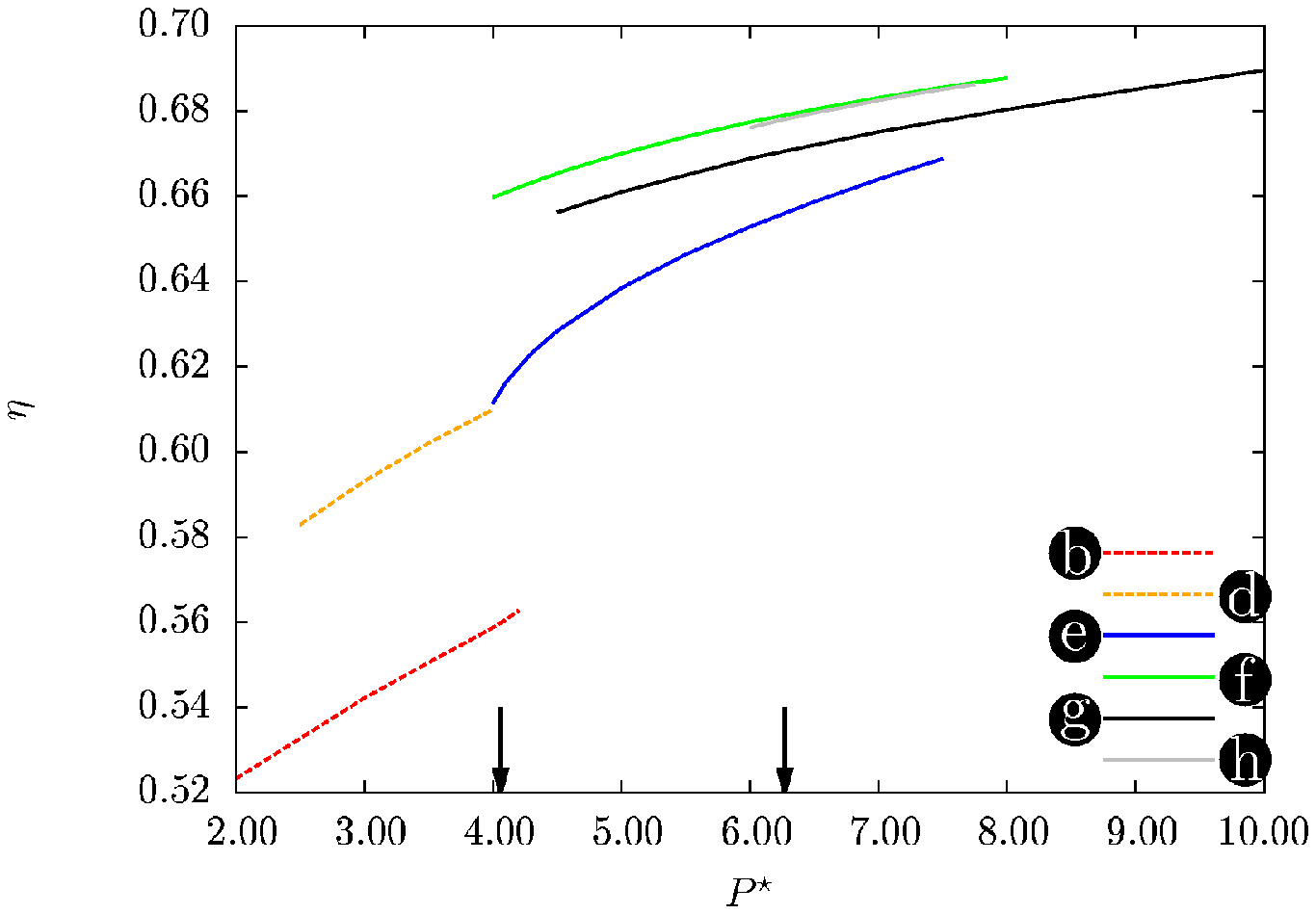}
\end{center}
  \caption{(Color online) Reduced binding energy (i.e., lattice sum)
    $U^{\star}$ (top panel) and packing fraction $\eta$ (bottom panel)
    as functions of $P^\star$ along the isotherm $T^\star = 0.10$. Results have been evaluated for the
    different crystal structures via MC simulations. The data are plotted 
    only over the ranges of mechanical stability of the respective structures.
    The vertical arrows indicate those pressure values where structural phase 
    transitions take place (i.e., between structures ``b'' and ``e'' at $P^\star=4.06$ and between structures ``e'' and ``f'' at $P^\star=6.27$).
    }
  \label{fig:isotherm_t0.1}
\end{figure}

In Figure \ref{fig:isotherm_t0.1} we have depicted $U^\star$ and
$\eta$ (both obtained via simulations) along the isotherm $T^\star =
0.10$ over a relatively large pressure interval, namely $P^\star \in
[2, 10]$. Values for the two quantities are depicted for structures
labeled ``b'', ``d'', ``e'', ``f'', ``g'', and ``h'' over the
respective ranges of stability; we note that mechanical stability
(i.e., a positive compressibility) is guaranteed as long as $\eta$
increases monotonically with increasing pressure. The corresponding
values for the lattice ``a'' coincide with the values of ``b'' within
line-thickness and lattice ``c'' is stable only for $T^\star \lesssim
0.006$ (see discussion above). As can be seen from the phase diagram
presented in \cite{patchy3d}, at this temperature the system forms at
low pressure values structure ``b'', which transforms at
$P^\star \simeq 4.06 $ into structure ``e'' and eventually at
$P^\star \simeq 6.27 $ into structure ``f''.

\begin{figure}[htbp]
\begin{center}
\includegraphics[width=9.cm]{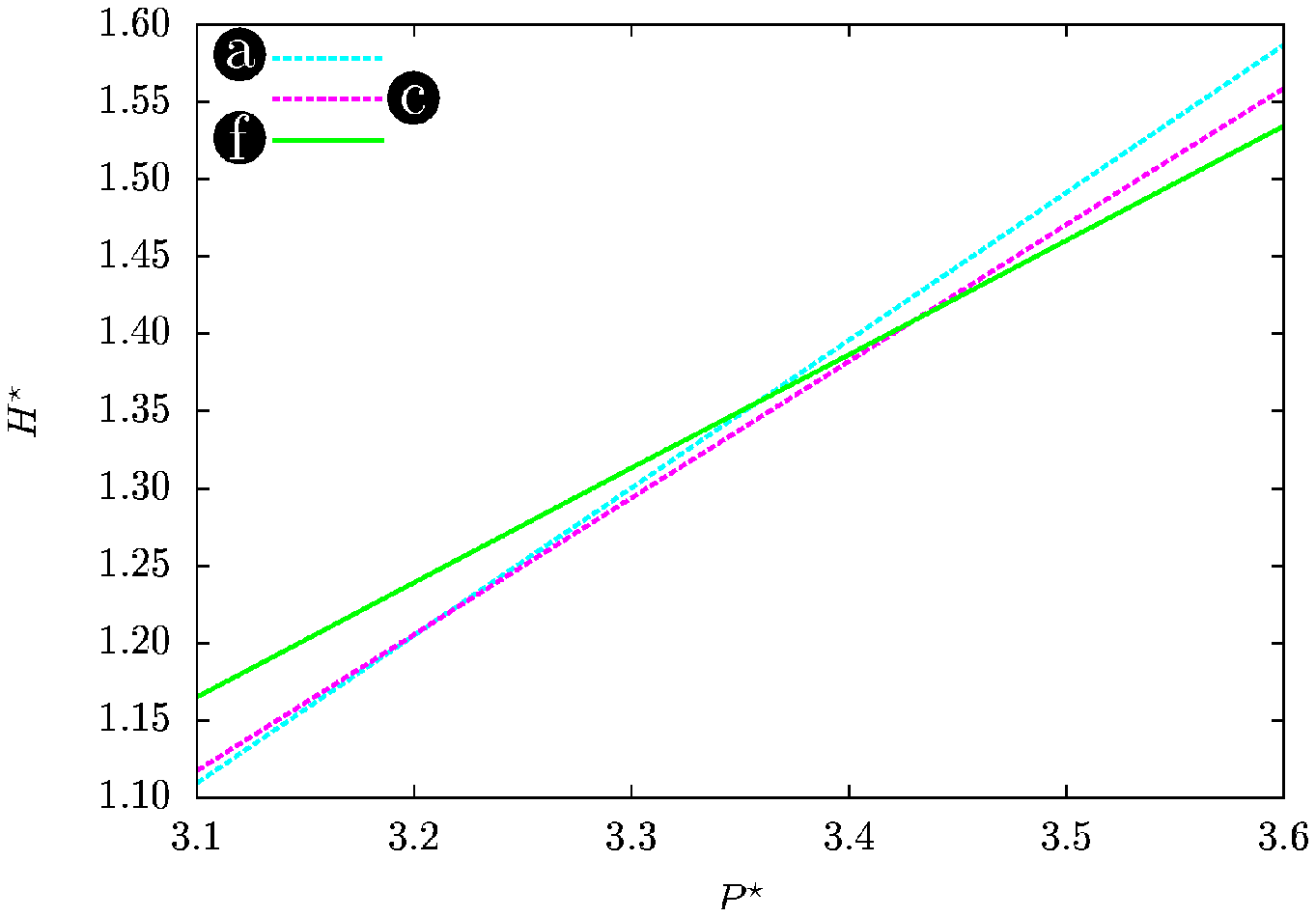}
\includegraphics[width=9.cm]{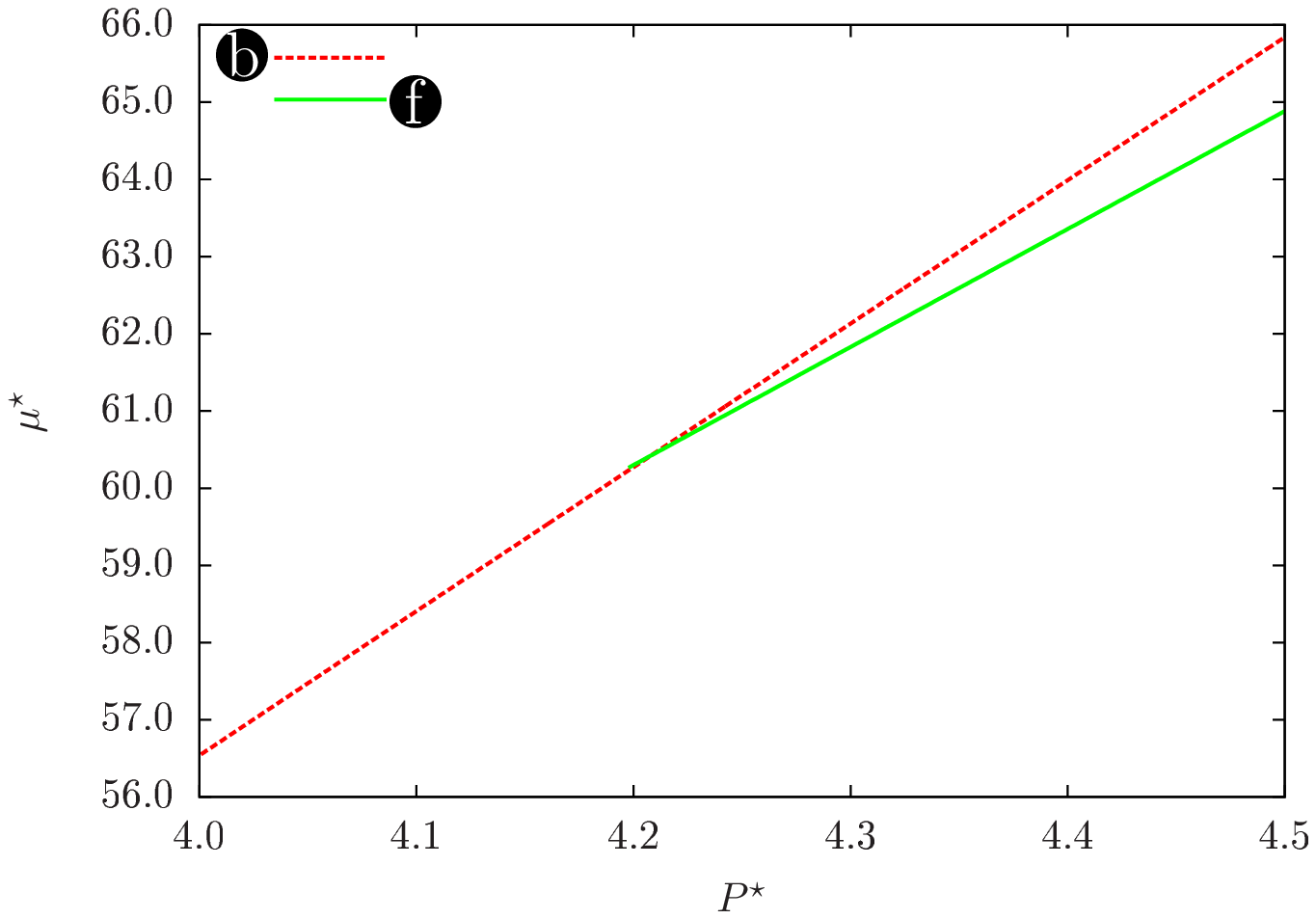}
\includegraphics[width=9.cm]{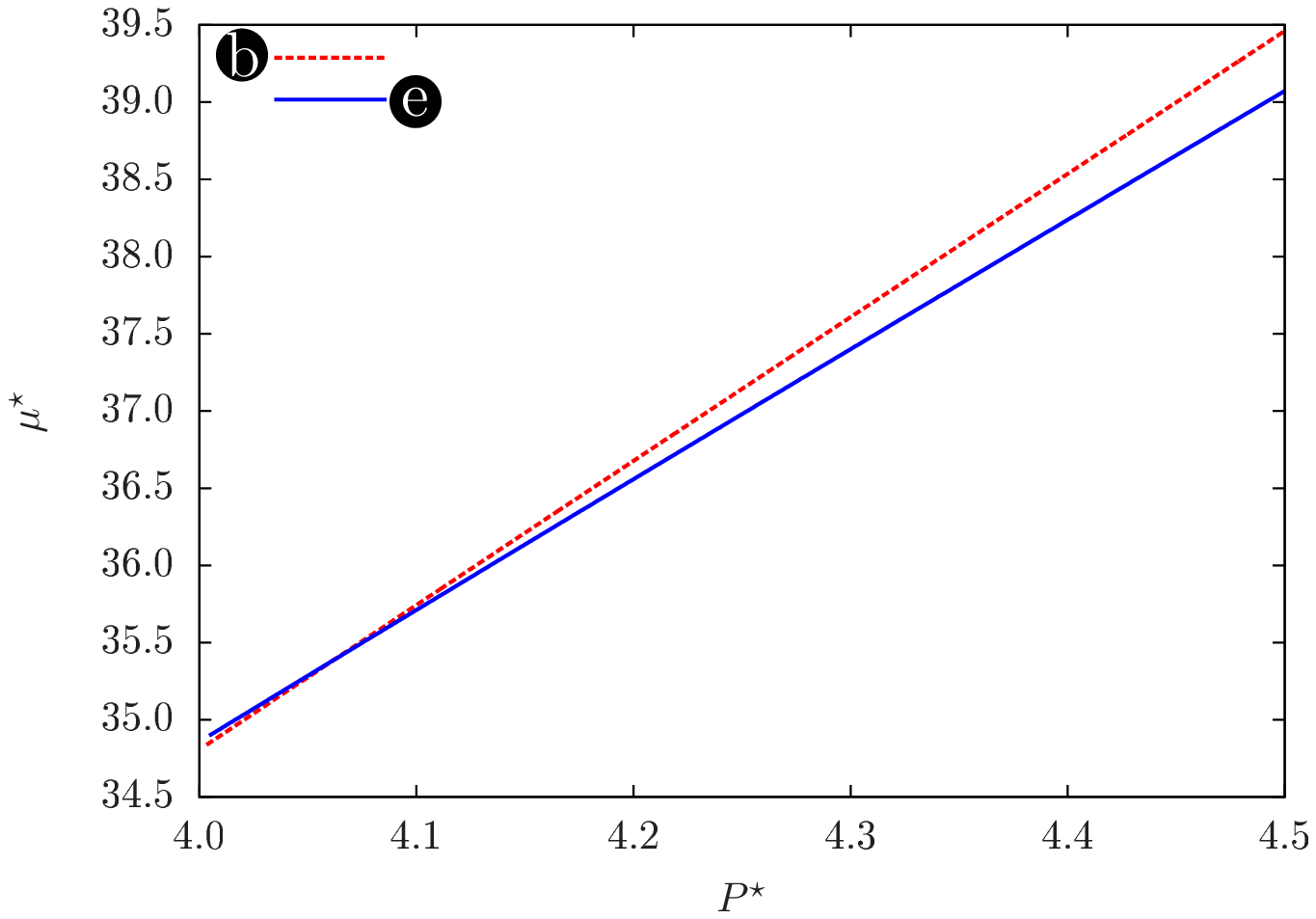}
\end{center}
  \caption{(Colour online) Top panel: reduced enthalpy $H^\star$ at
    vanishing $T^{\star}$ as a function of $P^\star$ in the vicinity
    of the phase transition between the ordered structures ``a'', ``c'', and ``f'' -- cf. phase
    diagram shown in \cite{patchy3d}. Center panel: reduced chemical
    potential $\mu^\star$ as a function of $P^\star$ along the
    isotherm $T^\star = 0.05$ in the vicinity of the phase transition
    between the ordered structures ``b'' and ``f''. Bottom
    panel: reduced chemical potential $\mu^\star$ as a function of
    $P^\star$ along the isotherm $T^\star = 0.10$ in the vicinity of
    the phase transition between the ordered structures ``b''
    and ``e''. The respective values in the vicinity of the
    phase transition between structures ``e'' and ``f'' are
    summarized in Table \ref{tab:trans_fcc-1-2}.}
\label{fig:enthalpy_mu}
\end{figure}

Finally, Figure \ref{fig:enthalpy_mu} shows the enthalpy $H^\star$ for
$T^\star = 0$ and the chemical potential $\mu^\star$ along two
different isotherms (i.e., $T^\star = 0.05$ and $T^\star = 0.10$) over
the narrow pressure ranges where a phase transition between two
competing structures takes place. The top panel shows for $T^\star = 0$ the enthalpy values in a
$P^\star$-range where the system transforms -- with increasing
pressure -- from a double diamond (``a'') to a double diamond/bcc
broken (``c'') lattice which eventually transforms into a
fcc-like structure (``f''). For
clarity the respective curves of the other structures are not
shown since they differ only by small amounts of the respective values
of the three energetically most favourable lattices. The
central panel of Figure \ref{fig:enthalpy_mu} focuses on the
transition region from structure ``b'' to lattice ``f'' along the
isotherm $T^\star = 0.05$. Again only minute differences in the
thermodynamic properties (here: the chemical potential) decide which
is the energetically most favourable ordered structure. Finally the
bottom panel of this Figure shows the chemical potential $\mu^\star$
along the isotherm $T^\star = 0.10$, focusing on the transition region
between structures ``b'' and ``e''. Again, the curves for
the other competing structures, characterized by (slightly) higher
chemical potential values are not shown for clarity. Table \ref{tab:trans_fcc-1-2} shows data for the chemical
potential for the phase transition between structures ``e'' and ``f'' at $T^\star=0.10$, where the differences in
the respective $\mu^\star$-values of the competing structures are too
small to be depicted in a figure. This indicates the strong
competition between the candidate structures.

\section{Conclusions}
\label{sec:Conclusions}

We have investigated the ordered equilibrium structures of patchy
particles with a regular tetrahedral patch decoration. To this end we
have identified at {\it vanishing} temperature with a reliable and
efficient optimization tool the ordered structures that correspond in
an NPT ensemble to the lowest-lying local enthalpy minima. With these
candidate structures at hand, we have calculated with a suitably
adapted Monte Carlo simulation technique, the thermodynamic properties
of these lattices at {\it finite} temperature via thermodynamic
integration. With this combination of two complementary approaches we
compensate for the respective limitations of the two methods: the
reliability of the optimization tool avoids -- by providing a
comprehensive set of candidate structures for the evaluation of the
phase diagram -- that possible equilibrium structures are simply
forgotten; the computer simulations allow an accurate evaluation of
the thermodynamic properties of the suggested candidate structures at
any finite temperature and any pressure value.

Based on the knowledge of the entire (pressure vs. temperature) phase
diagram \cite{patchy3d} we have studied the thermodynamic properties of these
candidate structures along selected isobars and isotherms in
detail. To this end we have split up the respective enthalpy values
into their relevant contributions, namely energy (i.e., the
lattice sum) and volume (i.e., the packing fraction). From these
detailed investigations it becomes obvious that the selection of the
energetically most favourable lattice structure at a given state point
is the result of a strong competition between the two above mentioned
contributions to the thermodynamic potential. Already at vanishing
temperature, unsaturated bonds of a given ordered structure (leading
to a higher value of the lattice sum) may be compensated by a dense
packing, leading to a lower enthalpy value, compared to a lattice with better
bond saturation but a smaller packing fraction. As temperature is
included, entropy enters the competition between these
two contributions and the situation becomes even more intricate.

In particular we learn that -- even though an ordered structure might
correspond only to a local minimum in enthalpy at vanishing
temperature -- this structure can turn out to be the energetically
most favourable lattice in some pressure- and temperature-range. Thus, it is necessary to consider such local enthalpy minima as possible candidate structures when evaluating the phase diagram of patchy particles.

\ack Financial support by the Austrian Science Foundation (FWF) under
project Nos. W004, M1170-N16 and P23910-N16 is gratefully acknowledged. E.G.N. gratefully acknowledges support from Grants Nos. FIS2010-15502 from the Direccion General de Investigacion and S2009/ESP-1691 (program MODELICO) form the Comunidad Autonoma de Madrid.

\eject

\eject

\newpage

\begin{table}[!h]
\begin{center}
\begin{tabular}{ c || c  | c }
$P^\star$ & \parbox[c]{3.2cm}{$\mu^\star$ (fcc-like 1 \includegraphics[width=0.4cm]{figs/figure1j.eps} )} & \parbox[c]{3.2cm}{$\mu^\star$ (fcc-like 2 \includegraphics[width=0.4cm]{figs/figure1l.eps} )} \\
\hline
6.00 & 51.29 & 51.37 \\
6.10 & 52.11 & 52.14 \\
6.20 & 52.90 & 52.92 \\
6.30 & 53.70 & 53.69 \\
6.40 & 54.49 & 54.47 \\
6.50 & 55.29 & 55.23 \\
6.60 & 56.08 & 55.98 \\
\hline

\end{tabular}
\end{center}
\caption{Addendum to Figure \ref{fig:enthalpy_mu}: Reduced chemical potential $\mu^{\star}$ at $T^{\star}=0.10$ of the structures ``e'' and ``f'' for $P^\star$-values in the vicinity of the phase transition between these structures. (The values are too close for a clear visual representation in a graph)}
 \label{tab:trans_fcc-1-2}
\end{table}


\begin{thebibliography}{10}
\bibitem{note1}
For double diamond/bcc-like structures, particles belonging to different
  non-interacting diamond sublattices appear in different colours. For
  structures consisting of hexagonal layers, particles located in different
  layers are coloured in an alternating pattern. For fcc-like structures, the
  particles located at the vertices of the cube and the particles at the
  centers of the faces of the cube are coloured red and yellow, respectively.
\end{thebibliography}
\end{document}